\documentstyle[12pt]{article}
\newcommand{\be}{\begin{equation}}
\newcommand{\ee}{\end{equation}}
\newcommand{\ba}{\begin{eqnarray}}
\newcommand{\ea}{\end{eqnarray}}

\newcommand{\bi}{\bibitem}
\begin{document}
\begin{center}
{\bf\Large
{Symmetries of the dual metrics}}
\end{center}
\begin{center}
Dumitru Baleanu\footnote[1]
{On leave of absence on
Institute of Space Sciences, P.O.Box, MG-23, R 76900,\\
Magurele-Bucharest, Romania,
baleanu@venus.ifa.ro}
\end{center}
\begin{center}
 Department of Mathematics and Computer Science,\\
Faculty of Arts and Sciences, Cankaya University,\\
 Ankara, 06350 Balgat ,Turkey, dumitru@cankaya.edu.tr
\end{center}

\begin{abstract}
In this paper the symmetries of the dual manifold were investigated.
We found the conditions when  the  manifold and its dual
admit the same Killing vectors and Killing-Yano tensors.
In the case of  an Einstein's metric $g_{\mu\nu}$ the corresponding
equations for its dual were found.
The examples of Kerr-Newman geometry and the separable
coordinates in $1+1$ dimensions were analyzed in details.
\end{abstract}
 
\section{Introduction}
 In a geometrical setting , symmetries are
connected with isometries associated with Killing vectors, and more
generally, with
 Killing tensors on the configuration space of the system.An
example  is the motion of a point particle in a space with isometries
which is a physicist's way of studying the geodesic structure of a manifold
\cite{hol1}.
Such studies were
extended to spinning space-times described by supersymmetric extensions
of the geodesic motion \cite{gib} and
it was shown that this
can give rise to interesting new types of supersymmetry as well.
The "non-generic" symmetries were investigated in the case of
Taub-NUT metric \cite{hol2} and extended Taub-NUT metric \cite{bal}.
It was a big
 success of Gibbons et al. \cite{gib} to have been able to show that the
Killing-Yano tensor \cite{kyano}, which had long been known to 
relativists
as a rather mysterious structure, can be understood as an object generating
a "non-generic" supersymmetry, i.e a supersymmetry appearing only in
specific space-times.
 Killing tensors are important for solving the
equations of motion in particular space-times.
The notable example here
is the Kerr metric which admits a second rank Killing tensor
\cite{gib}.

In \cite{hol3} a new geometric duality was introduced
but the physical interpretation of the dual metrics was not yet clarified.
 Let us consider a space with metric $g_{\mu\nu}$ admiting
 a Killing tensor field
 $K_{\mu\nu}$ \cite{hol3}.
The equation of motion of a particle on a geodesic is derived from the action
$S=\int{d\tau({1\over 2} g_{\mu\nu}\dot{x^{\mu}}{\dot x^{\nu}})}$.
The Hamiltonian is constructed in the form
 $H={1\over 2}g_{\mu\nu}p^{\mu}p^{\nu}$
and the Poisson brackets are $\{x_{\mu}, p^{\nu}\}=\delta_{\mu}^{\nu}$.
The equation of motion for a phase space function $F(x,p)$ can be computed
from the Poisson brackets with the Hamiltonian
${\dot F}=\{F,H\}$, where ${\dot F}={dF\over d\tau}$.
>From the covariant component $K_{\mu\nu}$ of the Killing tensor we
can construct a constant of motion $K$, with
$K={1\over 2}K_{\mu\nu}p^{\mu}p^{\nu}$.
It can be easily verified that
$\{H,K\}=0$.
The formal similarity between the constants of motion H and K ,
and the symmetrical  nature of the condition implying the existence
of the Killing tensor amount to a reciprocal relation between two
different models: the model with Hamiltonian H and constant of motion K,
and a model with constant of motion H and Hamiltonian K. The relation
between the two models has a geometrical interpretation: it implies that
if $K_{\mu\nu}$ are the contravariant components of a Killing tensor with
respect to the metric $g_{\mu\nu}$, then $g_{\mu\nu}$ must represent a
Killing tensor with respect to the metric defined by $K_{\mu\nu}$.
 When $K_{\mu\nu}$ has an inverse we interpret it as the
metric of another space and we can define the associated
Riemann-Christoffel connection $\hat\Gamma_{\mu\nu}^{\lambda}$ as usual
through the metric postulate ${\hat D}_{\lambda}K_{\mu\nu}=0$,
where ${\hat D}$ represents the covariant derivative with respect to
$K_{\mu\nu}$.

 Recently, Killing tensors of third rank in $1+1$ dimensional
geometry were classified \cite{ros} and the Lax
tensors on the dual metric were investigated \cite{dan}. The non-standard
Dirac operators which differ from, but commute with the standard Dirac
operator,
were analyzed on a manifold with  non-trivial Killing tensor admitting a
square root of Killing-Yano type \cite{holtena}.

 For these reasons  the symmetries of the dual manifolds are
  interesting to investigate.

The plan of this paper is as follows:\\
In Section 2  the  symmetries
of a dual manifold were investigated.
In Section 3  two examples were analyzed.
Our conclusions were presented in Section 4.

\section{ Dual metrics symmetries}

A Killing tensor is a
symmetric tensor which satisfies the following relation:
\be\label{kil}
D_{\lambda}K_{\mu\nu} +D_{\mu}K_{\nu\lambda} +D_{\nu}K_{\lambda\mu} = 0,
\ee
where D represent the covariant derivative with respect $g_{\mu\nu}$.

If the Killing tensor is non-degenerate  one can  find the
 associate Riemann-Christoffel of the dual manifold 
  \be\label{cristi}
{\hat\Gamma_{\mu\nu}^{\lambda}}=
{1\over 2}K^{\lambda\tau}({\partial K_{\mu\tau}\over\partial x^{\mu}}+
{{\partial K_{\nu\tau}\over\partial x^{\mu}}}
-{{\partial K_{\mu\nu}\over\partial x^{\tau}}}).
\ee
If the manifold  is torsion free,  taking into account
(\ref{kil}) and (\ref{cristi}), we found

\be\label{cone}
{\hat \Gamma^{\mu}_{\nu\lambda}}=\Gamma^{\mu}_{\nu\lambda}-
K^{\mu\delta}D_{\delta}K_{\nu\lambda},
\ee
   where
$\Gamma_{\mu\nu}^{\lambda}$  is
the associate Riemann-Christoffel connection with respect to $g_{\mu\nu}$.

We know that the conformal transformation of $g_{\mu\nu}$ is defined as
\ba\label{conf}
{\hat
g_{\mu\nu}}&=&e^{2U}g_{\mu\nu}\cr {\hat
g^{\mu\nu}}&=&e^{-2U}g^{\mu\nu}\qquad ,U=U(x).
\ea

Using (\ref{conf}) the  relation between the corresponding connections becomes
\be\label{game}
\hat\Gamma^{\lambda}_{\mu\nu}=\Gamma^{\lambda}_{\mu\nu}
 +2\delta^{\lambda}_{(\mu}U_{\nu)^{'}}- g_{\mu\nu}U^{'\lambda},
\ee
where $U^{'\lambda}={dU^{\lambda}\over dx}$.
 {}From (\ref{cone}) and (\ref{game}) we conclude  that the dual
transformation  (\ref{cone}) is not of a conformal transformation type.

   For this reason  it would be interesting  to investigate which are
the conditions
when the   manifold  admits the same Killing
vectors and Killing-Yano tensors as its dual manifold.

Let us denote by 
$\chi_{\mu}$  a Killing vector corresponding to $g_{\mu\nu}$ and
let $\hat\chi_{\mu}$ be a Killing vector corresponding to $K_{\mu\nu}$.

\vspace*{2mm}
{\bf Proposition 1}

The manifold  $g_{\mu\nu}$  and  its  dual
have the same
Killing vector iff
\be\label{vec}
 (D_{\delta}K_{\mu\nu})\hat\chi^{\delta}=0.
\ee

{\bf Proof.}

Since $\chi_{\sigma}$ is a Killing vector   we have
\be\label{killing}
D_{\mu}\chi_{\nu} +D_{\nu}\chi_{\mu}=0.
\ee

In the dual space using  (\ref{cone}) we found  the following
 corresponding equations

\be\label{kildual}
D_{\mu}\hat\chi_{\nu} +
D_{\nu}\hat\chi_{\mu}
+2K^{\delta\sigma}(D_{\delta}K_{\mu\nu}){\hat\chi_{\sigma}}=0,
\ee
where  D is
the covariant derivative on manifold $g_{\mu\nu}$  .\\
 If $\hat\chi_{\mu}=\chi_{\mu}$, then from (\ref{killing})
and
(\ref{kildual}) we get
\be
(D_{\delta}K_{\mu\nu})\hat\chi^{\delta}=0.
\ee
 Conversely, if (\ref{vec}) is satisfied, then from (\ref{kildual})
 we  deduce  immediately  $\chi_{\mu}=\hat\chi_{\mu}$.

q.e.d.
\vspace*{2mm}

A Killing-Yano $f_{\mu\nu}$ is an antisymmetric
tensor which satisfies the following equations \cite{kyano}
\be\label{yano1}
D_{\lambda}f_{\mu\nu} + D_{\mu}f_{\lambda\nu}=0.
\ee

 Let us suppose that a manifold  $g_{\mu\nu}$
has the Killing-Yano tensor $f_{\mu\nu}$ and its dual
admits the Killing-Yano tensor ${\hat f_{\mu\nu}}$.

\vspace*{2mm}
{\bf Proposition 2}

  The manifold  $g_{\mu\nu}$ and its dual  have the
same Killing-Yano tensor iff\be\label{cond}
{\hat f_{\nu\delta}}K^{\delta\sigma}D_{\sigma}K_{\mu\lambda} +
2{\hat f_{\sigma\lambda}}K^{\sigma\delta}D_{\delta}K_{\mu\nu} +
{\hat f_{\mu\sigma}}K^{\sigma\delta}D_{\delta}K_{\nu\lambda} =0.
\ee

{\bf Proof.}

Taking into account (\ref{cone})  Killing-Yano equations in the dual space
become
\ba\label{gendual}
 {\hat D_{\mu}}{\hat f_{\nu\lambda}} +{\hat
D_{\nu}}{\hat f_{\mu\lambda}}&=& D_{\mu} {\hat f_{\nu\lambda}} + D_{\nu}
{\hat f_{\mu\lambda}} + {\hat f_{\nu\delta}}K^{\delta\sigma}D_{\sigma}K_{\mu\lambda}
+ 2 {\hat f_{\sigma\lambda}}K^{\sigma\delta}D_{\delta}K_{\mu\nu}\cr & +&
{\hat f_{\mu\sigma}}K^{\sigma\delta}D_{\delta}K_{\nu\lambda} =0.
\ea

If
${\hat f_{\mu\nu}}=f_{\mu\nu}$,
using (\ref{gendual}) and because  $f_{\mu\nu}$ is a
Killing-Yano tensor on $g_{\mu\nu}$ we conclude that (\ref{cond}) is
identically zero.

 Conversely, if (\ref{cond}) is satisfied then from (\ref{gendual}) and
(\ref{yano1}) we have
 $\hat f_{\mu\nu}= f_{\mu\nu}$

q.e.d.
\vspace*{2mm}

{\sl Remarks}

 i) The associated metric
\be\label{formula}
K_{\mu\nu}=f_{\mu\lambda}f_{\nu}^{\lambda}
\ee
 has the inverse if
$f_{\mu\lambda}$ is non-degenerate.

ii) If the manifold  admits two
non-degenerate Killing-Yano tensors $f_{\mu\nu}$ and $F_{\mu\nu}$
then the dual metric has the form
$K_{\mu\nu}=f_{\mu\lambda}F^{\lambda}_{\nu} + f_{\nu\lambda}F^{\lambda}_{\mu}$

An important question is if the dual metric  satisfies  the
Einstein's equations .
For this reason is interesting to calculate
the connection between Riemann
curvature tensor, Ricci tensor, Ricci scalar of the  manifold  $g_{\mu\nu}$
and the corresponding expressions of the dual manifold.\\
 We know that  \be
R^{\beta}_{\nu\rho\sigma}=\Gamma^{\beta}_{\nu\sigma,\rho}-\Gamma^{\beta}_{\nu\rho,\sigma}
+\Gamma^{\alpha}_{\nu\sigma}\Gamma^{\beta}_{\alpha\rho}-\Gamma^{\alpha}_{\nu\rho}\Gamma^{\beta}_{\alpha\sigma}.
\ee
 Using (\ref{cone}) the corresponding dual Riemann curvature tensor
$\hat R^{\beta}_{\nu\rho\sigma}$ has the following expression
\be
\hat R^{\beta}_{\nu\rho\sigma}=R^{\beta}_{\nu\rho\sigma}
+R^{'\beta}_{\nu\rho\sigma},
\ee
where $R^{'\beta}_{\nu\rho\sigma}$ is
\ba
R^{'\beta}_{\nu\rho\sigma}&=&
-{(K^{\beta\delta}D_{\delta}K_{\nu\sigma})_{'\rho}}+
{(K^{\beta\chi}D_{\chi}K_{\nu\rho})_{,\sigma}}
-\Gamma^{\alpha}_{\nu\sigma}K^{\beta\chi}D_{\chi}K_{\alpha\rho}\cr
&-&
\Gamma^{\beta}_{\alpha\rho}K^{\alpha\delta}D_{\delta}K_{\nu\sigma}
+ \Gamma^{\alpha}_{\nu\rho}K^{\beta\chi}D_{\chi}K_{\nu\rho}
+ \Gamma^{\beta}_{\alpha\sigma}K^{\alpha\delta}D_{\delta}K_{\nu\rho}\cr
&-&K^{\alpha\delta}D_{\delta}K_{\nu\sigma}K^{\beta\chi}D_{\chi}K_{\alpha\rho} +
K^{\alpha\delta}D_{\delta}K_{\nu\rho}K^{\beta\chi}D_{\chi}K_{\alpha\sigma}.
\ea
The explicit expression of the Ricci tensor becomes
\be\label{ricci}
R_{\mu\nu}=\Gamma^{\alpha}_{\mu\alpha,\nu}-\Gamma^{\alpha}_{\mu\nu,\alpha}-
\Gamma^{\alpha}_{\mu\nu}\Gamma^{\beta}_{\alpha\beta}+\Gamma^{\alpha}_{\mu\beta}\Gamma^{\beta}_{\nu\alpha},
\ee
then the dual Ricci tensor  becomes
\be\label{riccidu}
{\hat R_{\mu\nu}}=R_{\mu\nu} +R^{'}_{\mu\nu}.
\ee
Here
\ba\label{riccih}
R^{'}_{\mu\nu}&=&  -{(K^{\alpha\chi}D_{\chi}K_{\mu\alpha})_{'\nu}}+
{(K^{\alpha\chi}D_{\chi}K_{\mu\nu})_{,\alpha}}
-\Gamma^{\alpha}_{\mu\nu}K^{\beta\delta}D_{\delta}K_{\alpha\beta}\cr
&-&
\Gamma^{\alpha}_{\mu\beta}K^{\beta\delta}D_{\delta}K_{\nu\alpha}-
 \Gamma^{\beta}_{\alpha\beta}K^{\alpha\chi}D_{\chi}K_{\mu\nu}
-  \Gamma^{\beta}_{\nu\alpha}K^{\alpha\chi}D_{\chi}K_{\mu\beta}\cr
&-&
K^{\beta\delta}K^{\alpha\chi}D_{\chi}K_{\mu\nu}D_{\delta}K_{\alpha\beta}
+
K^{\alpha\chi}D_{\chi}K_{\mu\beta}K^{\beta\delta}D_{\delta}K_{\nu\alpha}.
 \ea
If the metric $g_{\mu\nu}$ satisfies Einstein's equations in vacuum
\be
R_{\mu\nu}-{1\over 2}g_{\mu\nu}R=0,
\ee
then the  dual metric $K_{\mu\nu}$  satisfies the following equation
\be\label{egeneral}
({\hat R_{\lambda\sigma}}
-R^{'}_{\lambda\sigma})(\delta_{\lambda\mu}\delta_{\sigma\nu}-{1\over
2}g_{\mu\nu}g^{\lambda\sigma})=0,
\ee
where ${\hat R_{\mu\nu}}$ and $R^{'}_{\mu\nu}$ are given by
(\ref{riccidu}) and (\ref{riccih}).
 Using Eq.(\ref{egeneral}) we found that $K_{\mu\nu}$ is not an Einstein's
metric.
  The similar results can be obtained also in the presence of
matter and cosmological constant.
On the other hand if $R_{\mu\nu}=0$ , taking into account
(\ref{riccidu}) we can deduce that $R^{'}_{\mu\nu}$ is non-zero.
If the manifold $g_{\mu\nu}$ has  constant
 scalar curvature then the scalar curvature of dual manifold is not
constant.\\
 In the case of the Euclidean flat space, for example, we have a
Killing-Yano tensor of order two
$f_{\mu\nu}=a_{\mu\nu\lambda}x^{\lambda}$  which generate a Killing
tensor  $K_{\mu\nu}$ as
$K_{\mu\nu}=a_{\mu\gamma\lambda}a_{\nu\sigma}^{\gamma}x^{\lambda}x^{\sigma}$.
Here $a_{\mu\nu\lambda}$ is a constant antisymmetric tensor. When
$f_{\mu\nu}$ is non-degenerate  we can construct a non-degenerate
Killing tensor .
Let us consider for example the case of dimension N=4 and 
let $a_{\mu\nu\lambda}$
 be a constant tensor having components $\pm1$. The non-degenerate
Killing-Yano tensor $f_{\mu\nu}$ has the following components
\be
f_{12}=z+t, f_{13}=-y+t,f_{14}=-y-z,f_{23}=x+t,f_{24}=x-z,f_{34}=x+y.
\ee
Using (\ref{formula})the corresponding non-degenerate Killing tensor has the components
\ba\label{metrica}
&K_{11}=&(z+t)^2 +(-y+t)^2 +(-y-z)^2,K_{22}=(-z-t)^2 +(x+t)^2 +(x-z)^2\cr
&K_{33}=&(y-t)^2 +(-x-t)^2 +(y+x)^2,K_{44}=(z+y)^2 +(-x+z)^2 +(-y-x)^2\cr
&K_{12}=&(z+t)(x+t)+(-y-z)(x-z),K_{13}=(z+t)(-x+t)+(-y-z)(x+y)\cr
&K_{14}=&(z+t)(-x+z)+(-y+t)(-x-y),K_{23}=(-z-t)(y-t)+(t-z)(x+y)\cr
&K_{24}=&(-z-t)(y+z)+(x+t)(-x-y),\cr
&K_{34}=&(y-t)(y+z)+(-x-t)(-x+z).
\ea
It can be easily verified that the metric (\ref{metrica})
 is not an Einstein's metric. It has the scalar curvature 

\be
R={R_{1}\over R_{2}},
\ee
where
\newpage
 \ba
&R_{1}&=42t^7 +9x^5z^2 -67xt^6 +88t^3y^2xz +24y^3x^2zt +24ty^3z^2x +22z^4x^3\cr
 &+& 9z^6x-12x^4z^3 +41z^3x^3t+8z^2t^2x^2y-10z^3t^2xy +14z^2x^3yt +24x^4yzt\cr
 & +& 52t^3yx^2z- 276t^3yxz^2 -136t^3z^4 -167t^4z^3 +46t^4z^2x +18zx^5t +12x^4z^2\cr
 & +& 22x^3z^2y^2 +12x^2z^2y^3 +9xz^2y^4 +6z^6t -86z^5t^2 -296zt^4yx-146t^5z^2\cr
 &-& 61t^6z +198xz^4t^2- 102x^4t^3-168x^2t^5-12x^2z^5 -102x^3t^4 +44yzt^5 -36yt^5x\cr
 & +& 30zt^5x +9t^2x^5 +18ty^4xz + 15ty^2x^2z^2 +47ty^2xz^3-270t^2y^2x^2z+105t^2y^2xz^2\cr
 & +& 44zx^3y^2t +36z^3tx^2y +23z^5tx+ 145t^3z^3x +9ty^4z^2 +9t^2y^4x-117t^2y^4z\cr
 & -& 4z^3x^2y^2t +4z^3x^3y+4z^4x^2y +22z^4xy^4+ 12z^5xy +96t^3x^3z-226t^3z^2\cr
 &-& 118t^4x^2z-93t^2x^4z + 207t^2x^3z^2-6tx^4z^2-6z^4x^2t-255z^3t^2x^2\cr
 &+&12t^2x^4y -144t^2z^4-56t^3yz^3+22t^2x^3y^2 +19z^4ty^2-261z^3t^2y^2-164t^3yx^3\cr
&-&168y^3t^2z^2-180y^3t^3x-12y^3t^3z-233t^3y^2z^2+2t^4y^2x-174t^4y^2z +12ty^3z^3\cr
 & +& 12y^3z^3x +12y^3t^2x^2 +10z^5yt+ 34z^24t^4y-292x^2t^3y^2
+76t^4x^2y +24z^4txy\cr
&-& 152t^2x^3yz-168y^3t^2xz+168t^4y^3-126t^3y^4-224t^5y^2+112yt^6
\ea
and


 \ba
&R_{2}&=(3z^2t^2-6zt^2x +3t^2x^2-2x^2yt +2z^3t +4ztxy-2z^2yt +2x^3t-\!2z^2xt-2ztx^2\cr
&+&3x^2y^2+2yz^3-\!8xz^3 +3z^2y^2\!-\!6xzy^2+3x^4-\!2x^2zy-\!2z^2xy-\!8x^3z+2x^3y+3z^4\cr
&+&10z^2x^2)(3x^2z+3tx^2+2yxz +2t^2x +2yxt-2xz^2+5zt^2+5z^2t +3z^3+2yz^2 \cr
&+& 3t^3+3y^2z-2t^2y +3ty^2)(z+t)t^3.
\ea

\section{Examples}
\subsection{Kerr-Newman geometry}
The Kerr-Newman geometry describes a charged spinning black hole; in a
standard choice of coordinates, the metric is given by the following line
element \cite{gib}
\begin{eqnarray}\label{kerr}
 ds^2=-\frac{\Delta}{\rho^2}\left[dt-a\sin^2(\theta)d\varphi\right]^2+
\frac{\sin(\theta)^2}{\rho^2}\left[(r^2+a^2)d\varphi -adt\right]^2
+\frac{\rho^2}{\Delta}dr^2+\rho^2d\theta^2.
\end{eqnarray}
       Here
\begin{eqnarray}
\Delta=r^2+a^2-2Mr+Q^2,
\rho^2=r^2+a^2\cos^2\theta.
\end{eqnarray}

with $Q$ the background electric charge, and $J=Ma$ the
total angular momentum. The expression for $ds^2$ only describes the fields
{\em outside} the horizon, which is located at
\begin{eqnarray}
r=M+(M^2-Q^2-a^2)^{1/2}.
\end{eqnarray}
The Killing- Yano tensor for the Kerr--Newman is defined by \cite{gib}
\begin{eqnarray}
\frac{1}{2}f_{\mu\nu}dx^{\mu}\wedge dx^{\nu}=\nonumber
a\,\cos\theta\, dr\wedge(dt-a\,\sin^2\theta\, d\phi)\nonumber\\
+r\,\sin\theta\, d\theta\wedge[-a\,dt+(r^2+a^2)\,d\phi].
\end{eqnarray}

The Kerr-Newman metric admits a second-rank Killing tensor field. It
can be described in this coordinate system by the quadratic form

\begin{eqnarray}\label{kerd}
dk^2=K_{\mu\nu}dx^{\mu}dx^{\nu}=\frac{a^2\cos^2\theta\Delta}{\rho^2}
\left[dt-a\sin^2\theta d\varphi\right]^2+ \nonumber \\
\frac{r^2\sin^2\theta}{\rho^2}
\left[(r^2+a^2)d\varphi -adt\right]^2
-\frac{\rho^2}{\Delta a^2\cos^2\theta}dr^2+\frac{\rho^2}{r^2}d\theta^2.
\end{eqnarray}

Kerr-Newman metric admits two Killing vectors $\partial\over\partial t$
and $\partial\over\partial\varphi$.
Using Proposition 1 we found  that
Kerr-Newman metric and its dual (\ref{kerd})
have the same Killing vectors.\\
Solving Killing-Yano equations
(\ref{gendual}) we found  no solution on the dual manifold.\\
The dual metric (\ref{kerd}) has scalar curvature

\ba
&R=&-2{{\cos^2\theta}r^4 -5r^4 +2\cos^2\theta Mr^3 +r^2a^2\cos^4\theta
 -r^2\cos^2\theta Q^2
-6a^2\cos^4\theta Mr \over {(r^2
+a^2\cos^2\theta)\cos^2\theta r^2}}\cr
& &-
{{{10a^2\cos^4\theta  Q^2}+10a^4\cos^4\theta}\over {(r^2
+a^2\cos^2\theta)\cos^2\theta r^2}}\ea

The existence of Killing-Yano of valence three is an interesting question
for Kerr-Newmann metric and its dual.

A tensor  $f_{\mu_1\mu_2\mu_3}$
is called a Killing-Yano tensor of valence $3$ if it
is totally antisymmetric and satisfies the equation
\begin{equation}\label{yah}
f_{\mu_1\mu_{2}\mu_{3};\lambda} + f_{\lambda\mu_{2}\mu_{3};\mu_{1}}= 0,
\end{equation}
where comma denotes the covariant derivative.

On the other hand we know that for every Killing-Yano tensor
$f_{\mu_1\mu_2\mu_3}$ we have new supersymmetric charges  (for more
details see Ref.\cite{vaman}).
In  our  case
we have four independent
components of $f_{\mu_1\mu_2\mu_3}$ namely, $f_{r\theta\phi},f_{r\theta t},
f_{\theta\varphi t},f_{r\varphi t}$ and 15 independent Killing-Yano  equations.
After some calculations we
found that (\ref{yah}), has no solution for (\ref{kerr}).
Solving dual Killing-Yano equations (\ref{gendual}) we found  no solution for
(\ref{kerd}).

\subsection{Separable coordinate systems in $1+1$-dimensional Minkowski space}
  Separable
orthogonal coordinate systems on $n$-dimensional manifolds
are characterized by St\"{a}ckel systems, which is a system of
$n$ linearly  independent Killing tensors $K_{ij}$ of order two,
including the metric tensor  \cite{benenti1}.
Including the Cartesian system there are 10 orthogonal coordinate systems
 in $1+1$-dimensional flat space such that the Klein-Gordon equation
may be separated. In all of them coordinate lines are either straight
lines or conic sections, the latter ones being arranged in one or two
confocal families. In the following list the curvilinear coordinates
are denoted by $\mu$ and $\nu$. We listed below those coordinates
systems we have used in this paper (for more details see  for example
Refs.\cite{franz2},\cite{eisen}).

1. Cartesian system: Coordinates $t$ and $x$.

2. Elliptic system: Elliptic coordinates $\mu$ and $\nu$ are defined by
\be
t^2=\mu\nu,\,  x^2=(1-\mu)(1-\nu),\qquad 0<\nu, \mu<1.
\ee
with $\mu$ and $\nu$ labeling ellipses of one and the same confocal family
with mutually orthogonal intersections, given by the equations
\be\label{eqelipt}
\frac{t^2}{\mu}+\frac{x^2}{1-\mu}=1.
\ee
  This system is defined in the square $|t| +|x|\leq 1$.

3. Hyperbolic system: Defined by the same equations as the elliptic ones,
 but with $1<\nu <\mu <\infty$, so that (\ref{eqelipt}) describes one family
of hyperbolas. This hyperbolic system which is a continuation of the elliptic one
to other domains in space-time, it may be defined in 4 wedges of space-time,
$|t|-|x|>1$ or  $|x|-|t|>1$.

Killing tensor $K_{ik}$ associated with this systems is constructed
from the flat metric in
elliptic (hyperbolic) coordinates
\be\label{asib}
ds^2=\frac{\mu-\nu}{4}\left(\frac{d\mu^2}{\mu(\mu-1)}-\frac{d\nu^2}{\nu(\nu-1)}\right).
\ee
 and  has the following  form  \cite{eisen}

\be\label{inverseKilling}
K_{ik}=\frac{\mu-\nu}{4}\left(\begin{array}{ccc}
\frac{1}{\mu(\nu+1)(\mu-1)} & 0 \\
0 & \frac{-1}{\nu(\nu-1)(\mu+1)} \\
\end{array}\right).
\ee

Taking into account (\ref{inverseKilling}) 
we found that  the dual metric has  the form

\be\label{inversa}
ds^2=\frac{\mu-\nu}{4}\left(\frac{d\mu^2}{\mu(\nu+1)(\mu-1)}
-\frac{d\nu^2}{\nu(\nu-1)(\mu+1)}\right).
\ee

Solving  Killing-Yano equations   corresponding to (\ref{asib}) we get

\be\label{yano}
f_{\mu\nu}=-\frac{{(\mu-\nu)}^{2}}{16{\mu\nu(\mu-1)(\nu-1)}}.
\ee

We found that dual Killing-Yano equations have  the following solution

\be\label{yanodual}
f_{\mu\nu}=-\frac{{(\mu-\nu)}^{2}}{16{\mu\nu(\mu-1)(\nu-1)(\nu +1)(\mu +1)}}.
\ee

\section{Conclusions}

In this paper the geometric duality between
local geometry described by
$g_{\mu\nu}$ and  the local geometry described by  Killing tensor $K_{\mu\nu}$
were investigated.
 We found  that the transformation (\ref{cone})
is not a conformal transformation.
The Killing vectors equations and the Killing-Yano equations were analyzed
on the dual manifold. When $D_{\lambda}K_{\mu\nu}=0$
the symmetries of manifold and its dual  coincide. We found the
equations satisfied by the  dual metric $K_{\mu\nu}$.
It was found that if $g_{\mu\nu}$ satisfies  Einstein's equations in
vacuum
the corresponding dual manifold $K_{\mu\nu}$ is not an Einstein's metric.
We have proved that the dual
Kerr-Newman metric admits the same Killing vectors  as Kerr-Newman
metric but it has no Killing-Yano tensor of order two and three.
In the case of the separable coordinates in $1+1$ dimensions the corresponding metric
and the dual metric have Killing-Yano tensors and the same Killing vectors.

\noindent The classification of all Riemannian manifolds admitting  a
non-degenerate Killing tensor is an interesting problem and it
is under investigation \cite{balnew}.

\section{Acknowledgments}
I am grateful to S.Manoff and N.Makhaldiani for useful
discussions.

\end{document}